\documentclass[12pt]{article}
\usepackage{amsmath,amssymb}

\usepackage{graphicx}
\usepackage{wrapfig}

\usepackage{hyperref}
\usepackage{cite}

\def\beq{\begin{eqnarray}}
\def\eeq{\end{eqnarray}}

\newcommand{\Tr}{\,\mathrm{Tr}\,}            % already defined

\newcommand{\be}{\begin{equation}}
\newcommand{\ee}{\end{equation}}
\newcommand{\bea}{\begin{eqnarray}}
\newcommand{\eea}{\end{eqnarray}}
\newcommand{\bg}{\begin{gather}}
\newcommand{\eg}{\end{gather}}
\newcommand{\bseq}{\begin{subequations}}
\newcommand{\eseq}{\end{subequations}}

\renewcommand{\ln}{\mathop{\rm ln}\nolimits}

\def\tr{\hbox{Tr}}

\def\be{\begin{eqnarray}}
\def\ee{\end{eqnarray}}
\def\lb{\label}

\begin{document}

%\title{\textbf{Entanglement  entropy  and the a-theorem}}
\title{\textbf{The a-theorem and entanglement entropy}}
%\title{\textbf{Entanglement  entropy  and 
%irreversibility \\
%of Renormalization Group flows}}

\vspace{1cm}
\author{ \textbf{
 Sergey N. Solodukhin$^\sharp$ }} %\copyright

\date{}
\maketitle

\begin{center}
  \hspace{-0mm}
	  \emph{ Laboratoire de Math\'ematiques et Physique Th\'eorique  CNRS-UMR
	7350 }\\
	  \emph{F\'ed\'eration Denis Poisson, Universit\'e Fran\c cois-Rabelais Tours,  }\\
	  \emph{Parc de Grandmont, 37200 Tours, France} \\
\end{center}

{\vspace{-11cm}
\begin{flushright}
%LMPT-TH/2010-***
\end{flushright}
\vspace{11cm}
}

%\hfill{\tt IUB-TH/***}\\\mbox{} \\
%\twocolumn[\hsize\textwidth\columnwidth\hsize\csname
%@twocolumnfalse\endcsname

%\maketitle \thispagestyle{empty}% \vspace*{.5cm}

\begin{abstract}
\noindent { The a-theorem is demonstrated for the RG flows of entanglement entropy  in two and four dimensions. In four dimensions we  relate it to the term quadratic in intrinsic derivative of the dilaton along the entangling surface in the dilaton  entropy. The a-theorem, similarly to the c-theorem in two dimensions,  then follows from the positivity of the 2-point function
of stress-energy tensor.  We suggest that the a-theorem, provided it is properly reformulated in terms of the entanglement entropy, may  follow from the c-theorem.
%We also make some remarks on the  possible extension of our results to six dimensions.
}
%\noindent {PACS: 04.70Dy, 04.60.Kz, 11.25.Hf }}
\end{abstract}
%\vskip 2.pc
%\maketitle

\vskip 2 cm
\noindent
\rule{7.7 cm}{.5 pt}\\
\noindent 
\noindent
\noindent ~~~$^{\sharp}$ {\footnotesize e-mail: Sergey.Solodukhin@lmpt.univ-tours.fr}

%\newpage
 %   \tableofcontents
%\pagebreak

\newpage

\section{ Introduction}
\setcounter{equation}0

An intriguing long-standing question is whether the Renormalization Group (RG)  flow in a unitary quantum field theory is irreversible.
Or, in a weaker form,  whether the two conformal field theories defined on the end points of the flow (UV and IR) are related by an irreversible RG evolution.
In two spacetime dimensions there is a complete answer to these questions. It was demonstrated in 1986 by Zamolodchikov \cite{Zamolodchikov:1986gt} that in two dimensions there exists a certain function (called c-function) which is monotonically decreasing
along the RG trajectories. At the fixed points of the RG flow, where the theory becomes conformal,  this function coincides with the central charge of 
the theory and, hence, the inequality $c_{UV}> c_{IR}$ follows. Moreover, the RG flow in two dimensions occurs to be  a gradient flow thus  forbidding the recurrent behaviors such as limit circles. 

This remarkable theorem   motivated  the search for an extension of  
 this result, or some part of it, to higher dimensions.     In two dimensions the central charge is related to the conformal anomaly $<T^\mu_{\ \mu}>=\frac{c}{24\pi}R$, where $R$ is
 the Ricci scalar. Namely this interpretation of $c$ can be easily generalized. 
 In particular, in four dimensions,  there are two contributions to the trace anomaly\footnote{We use the following conventions: $[\nabla_\mu,\nabla_\nu]V^\alpha=R^\alpha_{\ \beta\mu\nu}V^\beta$,
 $R_{\mu\nu}=R^\alpha_{\ \mu\alpha\nu}$.}
 \be
 <T^\mu_{\ \mu}>=\frac{1}{64}(-a E_4+b W^2)\, ,
 \lb{1}
 \ee
where $W^2$ is the square of the Weyl tensor and $E_4$ is the  Euler density
\be
&&W^2=R_{\alpha\beta\mu\nu}R^{\alpha\beta\mu\nu}-2R_{\mu\nu}R^{\mu\nu}+\frac{1}{3}R^2\, , \nonumber \\
&&E_4=R_{\alpha\beta\mu\nu}R^{\alpha\beta\mu\nu}-4R_{\mu\nu}R^{\mu\nu}+R^2\, .
\lb{2}
\ee
In 1988 Cardy suggested that in the 4d RG flows  the central charge $a$ plays  a  role similar to the 2d central charge $c$ and conjectured that
\be
a_{UV}>a_{IR}\, .
\lb{3}
\ee
Since then the focus has been made on finding a proof of  the a-theorem  (see for instance \cite{Cappelli:1990yc}, \cite{Osborn:1989td})
that occurred to be a difficult problem. 
An elegant  proof has been suggested only recently by Komargodski and Schwimmer \cite{Komargodski:2011vj} (see the subsequent discussion in 
\cite{Komargodski:2011xv}, \cite{Luty:2012ww}, \cite{KST}, \cite{Fortin:2012hn}, \cite{Elvang:2012yc}, \cite{Elvang:2012st}, \cite{Dvali:2012zc} and in a recent review  \cite{Nakayama:2013is}). The idea of \cite{Komargodski:2011vj} is to promote the mass parameters of operators that explicitly break the
conformal invariance to functions of spacetime, $M_i\rightarrow M_i e^{-\tau(x)}$, by introducing the dilaton field $\tau(x)$.  Provided the dilaton changes under the conformal transformations, $\tau\rightarrow \tau+\sigma$, this procedure restores the conformal invariance. The quantum effective action then becomes a functional of the background fields, the dilaton $\tau(x)$ and the metric $g_{\mu\nu}(x)$, and encodes the information on $n$-point  correlation functions of stress-energy tensor. The dependence
of the effective action on the dilaton is constrained by the conformal symmetry and by the requirement that the conformal anomaly is now produced both by conformal transformations of  metric and of the dilaton.  Comparing now the conformal anomaly in two CFT's, UV and IR, the difference $\Delta a=a_{UV}-a_{IR}$ is due to the dilaton. In flat spacetime
and in the low energy approximation the anomaly part of the  dilaton action is quartic in derivatives and is responsible for element of $2\rightarrow 2$ dilaton scattering.
The positivity of $\Delta a$ then follows from a positive-definite dispersion relation satisfied by the S-matrix elements.  

On the other hand, it has been suggested by many authors that the a-theorem can be approached by using the  entanglement entropy \cite{entanglement} (for reviews on entanglement entropy see \cite{entanglement:reviews}). Numerical analysis of \cite{Latorre:2004pk} and analytic study of \cite{Casini:2004bw} based on the  strong subadditivity 
demonstrate that the c-theorem in $(1+1)$-dimensional spacetime can be reformulated in terms of the entanglement entropy. Taking the similarity between the RG flow and the geometric Ricci flow we mention also  a similar result \cite{Solodukhin:2006ic} on the monotonic behavior of the entanglement entropy under the Ricci flow in two dimensions.
Further progress in this direction was made in three dimensions \cite{Klebanov:2012yf}, \cite{Klebanov:2011td} using the F-theorem and in higher dimensions using the holographic duality \cite{Myers:2010tj}, \cite{Casini:2011kv}, \cite{Casini:2012ei}.

In four dimensions the central charge $a$ appears in the logarithmic term of the entanglement entropy of  a sphere \cite{Solodukhin:2008dh} (see \cite{Casini:2010kt} for numerical analysis and extensions to higher dimensions). Moreover, the conformal anomaly in the entanglement entropy can be integrated \cite{Solodukhin:2011zr} and, for the anomaly produced by the central charge $a$, reduces to a purely two-dimensional anomaly defined on the co-dimension two entangling surface.  These observations suggest that the a-theorem, being reformulated in terms of the entanglement entropy, could be approached along the same lines and using similar methods  as the Zamolodchikov's  c-theorem. 

In this note we make a step in this direction. We apply the approach of Komargodski and Schwimmer to the entanglement entropy and identify the part of the entropy which depends on  the dilaton field $\tau(x)$.
This part comes from  a certain term in the dilaton effective action which describes the coupling of the Einstein tensor to $\partial_\mu\tau\partial_\nu\tau$. Since the dilaton couples
to the trace of stress-energy tensor $\int d^d x \, \tau(x) T(x)$, $T(x)=T^\mu_{\ \mu}(x)$, the specified term originates from the 2-point correlation function of the trace
$\left\langle T(x), T(y)\right\rangle$ considered on a curved background metric. We then find that  $\Delta a>0$ as a consequence of the  positive-definite property of the spectral representation of the two-point function. This is exactly the property which can be used, as explained in  \cite{Cappelli:1990yc} and \cite{Komargodski:2011xv}, to prove the c-theorem in two dimensions. 

The paper is organized as follows. In section 2 we give a brief discussion of the approach of Komargodski and Schwimmer. The conformal properties of the entanglement entropy
in four dimensions are reviewed in section 3.  In section 4 we discuss the RG flows of the entanglement entropy in a two-dimensional theory  and give yet another proof of the c-theorem.  The RG flows of the entanglement entropy in a four-dimensional field theory and the property (\ref{3}) are studied in section 5.   
In section 6 we suggest that the similarity between the a- and c- theorems may not be incidental and, in fact, the a-theorem 
may  follow from the c-theorem.
We conclude and summarize in section 7.

\section{ The a-theorem}
\setcounter{equation}0
As we already mentioned in the introduction the idea of Komargodski and Schwimmer is to introduce a new scalar field, the dilaton $\tau$. Under the conformal transformations
the metric $g_{\mu\nu}$ and the dilaton $\tau$ transform  as
\be
g_{\mu\nu}\rightarrow e^{2\sigma}g_{\mu\nu}\, , \, \, \tau\rightarrow \tau +\sigma\, 
\lb{2.1}
\ee
so that the combination $\hat{g}_{\mu\nu}=e^{-2\tau}g_{\mu\nu}$ is conformal invariant. The metric and the dilaton are considered as the background fields.
The effective action then is a functional of both, $W_{eff}[g,\tau]$, and is sum of a conformally invariant part $W_{inv}$ and a part which generates the anomaly
$W_{anom}$. The invariant part can be easily constructed as an expansion  in curvature of the rescaled metric $\hat{g}_{\mu\nu}$,
\be
W_{inv}=-\int d^4x \sqrt{\hat{g}}\left(\kappa_0\hat{R}+\kappa_1 \hat{R}^2+\kappa_2\hat{E_4}+\kappa_3\hat{W}^2+..\right)\, .
\lb{2.2}
\ee
Being considered on a flat background, $g_{\mu\nu}=\delta_{\mu\nu}$, this is an expansion in a number of derivatives of the dilaton $\tau$. To the lowest order the equation of motion for the dilaton comes from the first term in (\ref{2.2}),
\be
e^{-\tau}=1-\varphi\, , \ \ \Box \varphi=0 \, .
\lb{2.3}
\ee
The anomaly part of the effective action is obtained by integrating the  conformal anomaly of the type (\ref{1}). The result of the integration is\footnote{We are working in the Euclidean signature. Therefore the anomaly (\ref{1}) and the dilaton action differ by sign from those that appear in \cite{Komargodski:2011vj}, see also \cite{Nakayama:2013is} for the analysis in the Euclidean signature.} 
\be
&&W_{anom}=\frac{\Delta a}{64} \int d^4x\sqrt{g}\left(\tau E_4+4(R^{\mu\nu}-\frac{1}{2}g^{\mu\nu}R)\partial_\mu\tau\partial_\nu\tau-4(\partial\tau)^2\Box\tau+2(\partial\tau)^4\right)\nonumber \\
&&-\frac{\Delta b}{64}\int d^4x \sqrt{g}\tau W^2_{\alpha\beta\mu\nu}\, .
\lb{2.4}
\ee
By using the condition that  the anomalies in the UV and IR conformal field theories should match one obtains that $\Delta a=a_{UV}-a_{IR}$. In a flat space limit only the terms with four derivatives of the dilaton in (\ref{2.4}) survive. By using the low energy equation of motion (\ref{2.3}) for the dilaton one finds that
\be
W_{anom}=-\frac{\Delta a}{32} \int d^4 x(\partial\tau)^4\, .
\lb{2.5}
\ee
The action (\ref{2.5}) defines the 4-point scattering amplitude  ${\cal A}(s,t)$, where $s$ and $t$ are the Mandelstam variables,  for which one derives a dispersion relation \cite{Komargodski:2011vj} (see also the subsequent discussion in \cite{Luty:2012ww})
\be
\frac{\Delta a}{64}=\frac{1}{4\pi}\int_{s>0}\frac{Im {\cal A}(s)}{s^3}\, .
\lb{2.6}
\ee
The desired relation (\ref{3}) is obtained by noting that the right hand side in equation (\ref{2.6}) is a positive quantity in a unitary theory. 
Since the dilaton $\tau(x)$ couples  to the trace of the stress-energy tensor  $T(x)$ the amplitude ${\cal A}(s,t)$ can be expressed in terms of 4-point correlation function of $T(x)$.
Thus, in the approach of Komargodski and Schwimmer the a-theorem is due to a certain positive-definite relation extracted from the 4-point function.

\section{ Conformal properties of entanglement entropy}
\setcounter{equation}0
\subsection{Entanglement entropy in the replica method}
The entanglement entropy is defined by first taking a pure, typically vacuum, state and then tracing the modes which reside inside an entangling surface $\Sigma$.
The result is a mixed state with a certain density matrix $\rho$ with entropy $S=-\Tr\rho\ln\rho$. 
The calculation of the entanglement entropy   can be carried out by using the so-called conical singularity method. It consists in introducing a small conical singularity with angle deficit $\delta=2\pi (1-\alpha)$ in the two-dimensional sub-space orthogonal
to the entangling surface $\Sigma$ so that locally, in a small vicinity of $\Sigma$, the space-time ${\cal M}_\alpha$ looks like a direct product $C_{2,\alpha}\times \Sigma$ of two-dimensional conical space $C_{2,\alpha}$ and surface $\Sigma$.
We define
\be
-\ln\Tr\rho^\alpha =W(\alpha)\, 
\lb{3.1}
\ee 
as an effective action of the quantum field theory in question on the conical background. The entropy then is defined by differentiating the effective action with respect to
$\alpha$,
\be
S=(\alpha\partial_\alpha-1)W(\alpha)|_{\alpha=1}\, .
\lb{3.2}
\ee

\subsection{Geometric invariants of conical space}

In a  curved metric the effective action can be represented as an expansion in the curvature. In the presence of a conical singularity the curvature 
has a delta-like contribution at the singular surface \cite{Fursaev:1995ef}
\begin{eqnarray}
&&R^{\mu\nu}_{\ \ \alpha\beta} = \bar{R}^{\mu\nu}_{\ \
\alpha\beta}+
2\pi (1-\alpha) \left( (n^\mu n_\alpha)(n^\nu n_\beta)- (n^\mu
n_\beta)
(n^\nu n_\alpha) \right) \delta_\Sigma \, ,\nonumber \\
&&R^{\mu}_{ \ \nu} = \bar{R}^{\mu}_{ \ \nu}+2\pi(1-\alpha)(n^\mu
n_\nu)
\delta_\Sigma \, , \nonumber       \\
&&R = \bar{R}+4\pi(1-\alpha) \delta_\Sigma\, ,
\label{singular-curvature}
\end{eqnarray}
where
$n^\mu_k \, , \, k=1,\, 2$ are two orthonormal
vectors
orthogonal to the surface $\Sigma$, $(n_\mu n_\nu)=\sum_{k=1}^{2}n^k_\mu
n^k_\nu$,
and the quantity $\bar{\cal R}$ is the regular part of the curvature. Thus namely the  singular terms in the curvature (\ref{singular-curvature}) 
produce a contribution to the entropy (\ref{3.2}). 

Using  the relations (\ref{singular-curvature}) one may calculate the  integral of polynomials of curvature over conical space. 
It should be noted that the formulas (\ref{singular-curvature}) were obtained in \cite{Fursaev:1995ef}  under the assumption that  a rotational symmetry
is present in the plane orthogonal to the entangling surface. This assumption  is justified if $\Sigma$ is black hole horizon. It guarantees that the components of the extrinsic curvature of the surface $\Sigma$ vanish. However, in general, 
the rotational symmetry is not present. The typical examples are the cylinder or sphere in Minkowski spacetime. The extrinsic curvature is non-vanishing in this case and, thus, it should be taken into account. The contribution of the extrinsic curvature to the integral of the Euler density and of the square of the  Weyl tensor is known \cite{Solodukhin:2008dh} (in a purely geometric way this is systematically  studied in \cite{FPS}).
It can be summarized as follows
\be
\int_{{\cal M}_\alpha}E_4=\alpha\int_{{\cal M}_{\alpha=1}} E_4+8\pi (1-\alpha)\int_\Sigma R_\Sigma\, ,
\lb{3.3}
\ee
where $R_\Sigma$ is the intrinsic curvature of the entangling surface $\Sigma$, and for the integral of the square of the 
Weyl tensor
\be
\int_{{\cal M}_\alpha}W^2=\alpha\int_{{\cal M}_{\alpha=1}} W^2+8\pi (1-\alpha)\int_\Sigma K_\Sigma +O(1-\alpha)^2\, ,
\lb{3.4}
\ee
where we introduced the quantity
\be
K_\Sigma=R_{ijij}-R_{ii}+\frac{1}{
3}R -(\tr k^2-\frac{1}{ 2} k_i k_i )
\label{3.5} 
\ee 
where $R_{ijij}=R_{\alpha\beta\mu\nu} n^\alpha_i n^\beta_j n^\mu_i n^\nu_j$, $R_{ii}=R_{\alpha\beta} n^\alpha_i n^\beta_i$, and $k^i_{\mu\nu}$ is the extrinsic curvature of $\Sigma$ with respect to a normal vector $n^i$. 
The important point here is that the surface contribution due to the Euler density depends only on the intrinsic geometry of the surface $\Sigma$ while 
the surface contributions of other geometric quantities like the Weyl squared may depend on both the intrinsic and extrinsic geometry of the surface and on the geometry of the spacetime.

\subsection{Entanglement entropy and conformal anomaly}
Here we summarize the general relation of the entanglement entropy and the conformal anomalies in $d=4$ \cite{Solodukhin:2008dh}, \cite{Solodukhin:2011zr}.
Consider  a generic 4d conformal field theory, characterized by the conformal anomalies of type A and B. On a curved background (equipped with a metric $g_{\mu\nu}$)
with a singular surface $\Sigma$ (equipped with a 2d metric $\gamma_{ij}$) this theory  is described by the quantum effective action which has the bulk and boundary components. To first order in $(1-\alpha)$ one has that
\be
W=\alpha W_{bulk}+(1-\alpha){ W}_\Sigma \, .
\lb{18}
\ee
The entanglement entropy is determined by the surface part of the effective action,
\be
S=-W_\Sigma\, .
\lb{18-1}
\ee
The bulk part of the action has a standard decomposition in terms of the UV cut-off $\epsilon$ 
\be
W_{bulk}=\int_{M_{\alpha=1}} \left(\frac{a_4}{ \epsilon^4}+\frac{a_1}{ \epsilon^2}-a_2\ln\epsilon\right) +w(g)\,  .
\lb{19}
\ee
Under the conformal transformations of the bulk metric,  $g\rightarrow e^{2\sigma(x)} g$, the UV finite part $w(g)$ produces the anomaly
\be
w( e^{2\sigma} g) =w(g)+\int_M a_2\sigma(x)+O(\sigma^2)\, .
\lb{20}
\ee
In four dimensions one has that (see \cite{Duff:1993wm} for a review)
\be 
a_2=\frac{1}{64}(a E_{(4)}-b W^2)\, .
\label{21} 
\ee

Similarly, the surface term in the effective action (\ref{18}) is decomposed on the UV divergent and UV finite parts,
\be
{ W}_\Sigma=-\int_\Sigma\left(\frac{N}{ 48\pi\epsilon^2}+s_0\ln\epsilon\right) -s(g,\gamma)\, ,
\lb{22}
\ee 
where $N=N(A,B)$ is the effective number of fields in the theory, the fermions are counted with the weight $1/2$.
Under the conformal (Weyl) transformations of the  bulk metric and of the surface metric, $g\rightarrow e^{2\sigma} g$, $\gamma\rightarrow e^{2\sigma} \gamma$ the UV finite part $s(g, \gamma)$ transforms as
\be
s(e^{2\sigma} g,e^{2\sigma}\gamma)=s(g,\gamma)-\int_\Sigma s_0 \sigma +O(\sigma^2)\, .
\lb{23}
\ee 
The term $s_0$ is the surface part of the conformal anomaly  \cite{Solodukhin:2008dh},
\be
&&s_0=a s_A-bs_B \,, \nonumber \\
 &&s_A=
\frac{\pi}{ 8}R_\Sigma \, , \ \ s_B= \frac{\pi}{ 8}K_\Sigma\, ,
\label{24} 
\ee 
Under the conformal transformations the intrinsic curvature $R_\Sigma$ and the quantity $K_\Sigma$ transform as follows
\be
&&R_\Sigma(e^{2\sigma} \gamma)=e^{-2\sigma}(R_\Sigma(\gamma)-2\Delta_\Sigma\sigma)) \nonumber \\
&&K_\Sigma(e^{2\sigma} g,e^{2\sigma}\gamma))=e^{-2\sigma}K_\Sigma(g,\gamma)\, .
\lb{27}
\ee
The surface integral of $s_A$ is topological invariant, the Euler number of the surface. On the other hand, the surface integral of $K_\Sigma$
 is conformal invariant. 
 With the help of (\ref{27}) one can integrate (\ref{23}) and obtain the non-local  surface action and the entropy \cite{Solodukhin:2011zr}
\be
S=s(g,\gamma)=\frac{a\pi}{ 32}\int_\Sigma R_\Sigma \frac{1}{ \Delta_\Sigma} R_\Sigma -\frac{b\pi}{ 16}\int_\Sigma K_\Sigma \frac{1}{ \Delta_\Sigma} R_\Sigma +s_{conf}
\, ,
\lb{28}
\ee
where the last  term is a conformally invariant part and we skip the UV divergent terms.

To summarize the $a$-anomaly  produces a purely intrinsic contribution to the entanglement entropy. This is not so for the anomaly due to the Weyl squared.
There are certain cases, when $K_\Sigma=0$ and the respective contribution to the entropy is eliminated. The typical example is  when $\Sigma$ is a 2-sphere in Mikowski spacetime.
In this case the finite part of the entropy is purely intrinsic in the sense that it is entirely due to the intrinsic geometry of the entangling surface.

\subsection{The dilaton  entropy}

In the approach of Komargodski and Schwimmer the dilaton $\tau$ is yet another background field, such  as metric $g_{\mu\nu}$. Thus, the entanglement entropy
of the quantum field theory becomes a functional of the dilaton and of the metric. How the entropy depends on the  dilaton can be easily obtained by applying  
the method of conical singularity to the actions (\ref{2.2}) and (\ref{2.4}).  The conformally invariant part of the entropy then takes the following form\footnote{One might worry whether the $\hat{R}^2$ term in the action (\ref{2.2}) may give, in flat space, an extra contribution to the entropy that is entirely due to the extrinsic curvature of the entangling surface. This does not happen as is shown in the forthcoming paper \cite{FPS}.}
\be
S_{inv}=4\pi\int_\Sigma \sqrt{\hat{\gamma}}\left(\kappa_0+2\kappa_1\hat{R}+2\kappa_2\hat{R}_{\Sigma}+2\kappa_3\hat{K}_\Sigma+..\right)\, ,
\lb{4.1}
\ee
where $\hat{\gamma}=e^{-2\tau}\gamma$. The anomaly part of the entropy is obtained by applying (\ref{singular-curvature}) to the anomaly action (\ref{2.4}),
\be
S_{anom}=\Delta a\frac{\pi}{8}\int_\Sigma \sqrt{\gamma}\left(-\tau R_\Sigma (\gamma)+(\partial_\Sigma\tau)^2\right)+\Delta b\frac{\pi}{8}\int_\Sigma \sqrt{\gamma}\tau K_\Sigma\, .
\lb{4.2}
\ee
Here the term $(\partial_\Sigma\tau)^2=\gamma^{\mu\nu}\partial_\mu\tau\partial_\nu\tau$, where $\gamma^{\mu\nu}=g^{\mu\nu}-(n^\mu n^\nu)$ is the metric on the surface, is defined intrinsically on the surface $\Sigma$. Thus, the a-anomaly part in the dilaton action is purely intrinsic. This is consistent with what we said above regarding the a-anomaly  (\ref{28}).

In order to eliminate almost all terms in the entropy let us consider Minkowski spacetime and the entangling surface  $\Sigma$ to be  an infinite plane.
The entropy then contains only two terms
\be
S(\tau)=\Delta a\frac{ \pi}{8}\int_\Sigma (\partial_\Sigma \tau)^2+4\pi\kappa_0\int_\Sigma \sqrt{\gamma}e^{-2\tau}\, .
\lb{4.3}
\ee
We used the low energy equation of motion (\ref{2.3}) in the bulk  when derived (\ref{4.3})  so that the term $\hat{R}$ in (\ref{4.1}) does not produce any contribution. 
In what follows we shall focus on the first term in (\ref{4.3}) and ignore the second,   ``cosmological constant'', term. 

First of all we notice that the four-derivatives  terms  in (\ref{2.4}), which were very important in the analysis of Komargodski and Schwimmer, do not contribute at all
to the entanglement entropy! Instead, we have a contribution from the term with two derivatives of $\tau$ that couples to the Einstein tensor $G^{\mu\nu}=R^{\mu\nu}-\frac{1}{2}g^{\mu\nu}R$.
This term compensates in (\ref{2.4}) the conformal transformation of the Euler density, $\delta_\sigma E_4=8G^{\mu\nu}\nabla_\mu\nabla_\nu\sigma$.
Namely the coupling to the Einstein tensor was crucial in  establishing the fact that only the intrinsic derivatives of the dilaton along the surface appear in the entropy (\ref{4.3}).
The other remark is that the anomaly term in the entropy (\ref{4.3}) is produced by a term quadratic in $\tau$ in the dilaton action (\ref{2.4}). Since the dilaton $\tau$ couples to the trace $T(x)$ of the stress-energy tensor the anomaly term in the dilaton entropy   is related to a 2-point function of the trace $T(x)$, as we  show below.

\section{Analysis of the entanglement entropy in D=2}
\setcounter{equation}0
\subsection{Dilaton anomaly action and the entropy}
Let us start our analysis with a brief consideration of the two-dimensional case.  In two dimensions the anomaly part of the dilation action is \cite{Komargodski:2011xv}
\be
W_{anomaly}(\tau)=-\frac{\Delta c}{24\pi}\int(\tau R-(\nabla\tau)^2)\, ,
\lb{5.1}
\ee
where $\Delta c=c_{UV}-c_{IR}$ is the variation in the central charge of the theory in UV and IR regimes. Applying (\ref{singular-curvature}) to the dilaton action
(\ref{5.1}) we find the dilaton contribution to the entanglement entropy in $D=2$,
\be
S(\tau)=\frac{\Delta c}{6}\tau_\Sigma\, ,
\lb{5.2}
\ee
where $\Sigma$ is the discreet set of points of separation between the two subsystems and $\tau_\Sigma$ is the value of the dilaton on this set.
For simplicity we shall consider the case  when $\Sigma$ is just a single point separating two subsystems  of a quantum field defined on infinite line.

\subsection{A quick analysis of the dilaton entropy}
In order to illustrate the calculation of  $\Delta c$ let us consider a single  free massless field with the action $W_0(\phi)=\frac{1}{2}\int (\nabla\phi)^2$ and  perturb it  by a mass term $\frac{m^2}{2}\int \phi^2$. The dilaton is introduced by  replacing $m^2\rightarrow e^{-2\tau}m^2$. A quick, and not rigorous, way to find the dependence of the
entanglement entropy on the dilaton is the following. We know that in  the presence of mass $m$ the entanglement entropy  of a 2d scalar field is
\be
S=\frac{1}{6}\ln \frac{1}{\epsilon m}\, ,
\lb{5.3}
\ee
where $\epsilon$ is a UV cut-off. Now, replacing here $m\rightarrow m e^{-\tau_\Sigma} $ we find 
\be
S(\tau)=\frac{1}{6}\tau_\Sigma\,
\lb{5.4}
\ee
for the dilaton part in the entropy. That the value of the dilaton $\tau$ has to be taken at the point of separation $\Sigma$ is not obvious in this
quick derivation. However, the distributional nature of the conical singularity, which  is important in the derivation of the entanglement entropy,   tells us that the 
all quantities which vary in space  should be taken at the tip of the singularity, i.e. at $\Sigma$.

\subsection{Entanglement entropy of a free field}
Let us consider a slower calculation of the dilaton entropy (\ref{5.4}).
For small $\tau$ the perturbed action is 
 $W_{pert}(\phi)=W_m(\phi)-\int \tau (x) {\cal O}(x)$,
where $W_m(\phi)=W_0(\phi)+\frac{m^2}{2}\int \phi^2(x)$ is the action of a massive field and  ${\cal O}=m^2\phi^2$ is a perturbation. Now, to linear order in $\tau$, we have that 
\be
\left\langle e^{\int \tau(x) {\cal O}(x)}\right\rangle=e^{-\int W(\tau)}\, , \ \ W(\tau)=-\int \tau(x) \left\langle {\cal O}(x)\right\rangle \, ,
\lb{5.5}
\ee
where the expectation values are calculated in the massive theory. For operator ${\cal O}=m^2\phi^2$ the expectation value is expressed in terms of coincident points limit of
Green's function of  the massive field,
\be
{\cal O}(x)=m^2G_m(x,x)\, .
\lb{5.6}
\ee
In 2d Minkowski spacetime, and in the coincident points limit, Green's function is trivial, $G_m(x,x)=G_m(0)$. This is not so in the presence of a conical singularity. 

We shall use a representation for Green's function that uses the heat kernel,
\be
G_m(x,y)=\int_0^\infty ds \, K(s,x,y) e^{-m^2s}\, ,
\lb{5.7}
\ee
where $K(s,x,y)$ is the heat kernel of a massless field. In 2d Minkowski spacetime the heat kernel is
$$K_0(s,x,y)=\frac{1}{4\pi s}  e^{-\frac{(x-y)^2}{4s}}\, .
$$
Let us choose a polar coordinate system $(r,\theta)$ with the center  in the point $\Sigma$.
In the presence of a conical singularity one uses the Sommerfeld formula to make the heat kernel $2\pi\alpha$ periodic in $\theta$,
\be
K_\alpha(s,x,y)=K_0(s,x,y)+ \frac{i}{4\pi\alpha}\int_\Gamma \, \cot \frac{w}{2\alpha}\, K(s,x(w),y)\, dw \,  .
\lb{5.8}
\ee
The contour $\Gamma$
consists of two vertical lines, going from $(-\pi+ i \infty )$
to $(-\pi- i \infty )$ and from $(\pi- i \infty )$ to
$(\pi-+ i \infty )$ and intersecting the real axis between the
poles of the $\cot \frac{w}{2\alpha}$: $-2\pi\alpha$, $0$ and $0,$
$+2\pi\alpha$, respectively.
The second term in (\ref{5.8}) vanishes when $\alpha=1$. Namely this term in the heat kernel is important for the entropy calculation.
Let us expand the dilaton $\tau$ in powers of the radial coordinate $r$.  At the coincident points we find
$$
K(s,x(w),x)=\frac{1}{4\pi s}e^{-\frac{r^2\sin^2w/2}{s}}\, .
$$
The regularity of the dilaton in the Cartesian coordinates $(x_1,x_2)$ means that it can be decomposed in positive powers  of $x_1$ and $x_2$,  the monomes  $x_1^l x_2^{n-l}$ then form a basis of the decomposition. This can be translated to a decomposition in the polar coordinates,
\be
\tau(r,\theta)=\sum_{n=0}^\infty r^n\sum_{l=-n}^{l=n}\tau_{n,l} \, e^{il\theta/\alpha}\, ,
\lb{5.9}
\ee
where  $\tau_{0,0}=\tau_\Sigma$, is the value of the dilaton at $\Sigma$. We then find (see also \cite{Fursaev:1993qk} for a related calculation)
\be
\int_{{\cal M}_{\alpha}} \tau (x)  K_\alpha(s,x,x)= \frac{\alpha}{4\pi s}\int_{{\cal M}_{\alpha=1}}\tau(x) +\int\frac{1}{4}\sum_{k=0}^\infty \frac{k!}{(2k)!}\, \alpha\, P_{2k+2}(\alpha) \tau_{2k,0}\, s^k \, ,
\lb{5.10}
\ee
where we performed the integration over the radial coordinate $r$ and the angular coordinate $\theta$, and introduced
\be
\frac{i}{4\pi\alpha}\int_\Gamma dw \, \cot \frac{w}{2\alpha}\, \frac{1}{(\sin\frac{w}{2})^{n}}=P_n(\alpha)\, .
\lb{5.11}
\ee
It occurs that for odd values of $n$ the integral in (\ref{5.11}) vanishes. That is why in (\ref{5.10}) only even values of $n$ appear.
All functions $P_n(\alpha)$ vanish if $\alpha=1$. For $n=2$ and $n=4$ we find an exact form
\be
P_2=\frac{1}{3\alpha^2}(1-\alpha^2)\, , \ \ P_4(\alpha)=\frac{1}{45\alpha^3}(1-\alpha^2)(11\alpha^2+1)\, .
\lb{5.12}
\ee
Multiplying now (\ref{5.10}) by $e^{-m^2s}$ and integrating over the proper time $s$ we find the dilaton action
\be
W_\alpha(\tau)=\alpha W_{\alpha=1}(\tau)- \sum_{k=0}^\infty \frac{1}{4m^{2k}}\frac{(k!)^2}{(2k)!}\, \alpha\, P_{2k+2}(\alpha) \tau_{2k,0}\, .
\lb{5.13}
\ee
Applying now (\ref{3.2}) to (\ref{5.13}) we obtain the dilaton part in the entanglement entropy,
\be
S(\tau)=\frac{1}{6}\tau_\Sigma+O(\frac{1}{m^2}\partial_r^{2}\tau|_\Sigma)\, .
\lb{5.14}
\ee
We used here that $\tau_{0,0}=\tau_\Sigma$ and that $\tau_{2k,0}\sim \partial_r^{2k}\tau|_\Sigma$. The mass independent, no-derivative, term in the entropy (\ref{5.14}) is precisely
(\ref{5.2}), the term we are mostly interested in  in the context of the c-theorem. For a free field we the have that $\Delta c=1$ and, hence, $c_{IR}=0$, i.e. the IR theory does not contain any degrees of freedom.

\subsection{Entanglement entropy in general case, the c-theorem}
This free field calculation, however,  is useful in the study  of the RG flows in the general case. Indeed,  the expectation value of the operator $\cal O$ which couples to the dilaton can be represented as follows
\be
\left\langle {\cal O}\right\rangle=\lim_{x\rightarrow x'} \int dk\, {\cal O}(k)e^{-ik(x-x')}=\lim_{x\rightarrow x'} \int_0^\infty dm \, m^2 C(m)G_m(x,x')\, ,
\lb{5.15}
\ee
where for ${\cal O}(k)$ we have used a spectral representation
\be
{\cal O}(k)=\int_0^\infty dm\,  m^2 \frac{C(m)}{k^2+m^2}\, .
\lb{5.16}
\ee
If ${\cal O}$ is a positive operator, as it is in our case since $\cal O$ is the trace of the stress-energy tensor, then $C(m)\ge 0$.
The calculation in the general case (\ref{5.15}), thus, reduces to a free field calculation for the operator (\ref{5.6}). Repeating now the same steps as above, this time  for the operator (\ref{5.15}),  and focusing only on the no-derivative term in the dilaton entropy we find that
\be
S(\tau)=\frac{1}{6}\int_0^\infty dm\,  C(m)\, \tau_\Sigma\, .
\lb{5.17}
\ee
Comparing this to (\ref{5.2}) we find 
\be
\Delta c=\int_0^\infty dm \, C(m)\geq 0\, 
\lb{5.18}
\ee
and, thus, reproduce the c-theorem.

\section{Analysis of the dilaton action and  entanglement entropy in D=4}
\setcounter{equation}0
In this section we prefer to use the field $\varphi$ (\ref{2.3}) to represent the dilaton, for small values we have $\varphi=\tau$.
Our strategy here is to look at   the coupling $G^{\mu\nu}\partial_\mu\varphi\partial_\nu\varphi$ in the effective dilaton action.
Then, using the formulae (\ref{singular-curvature}) for the curvature due to a conical singularity we compute the corresponding contribution to the dilaton entropy.

\subsection{A free scalar field: the dilaton effective action and the entropy}
Let us start with a conformal scalar field and perturb it with a massive term.  The perturbed action then is
\be
W=\frac{1}{2}\int \phi\left(-\Box+\frac{1}{6}R+m^2+V(\varphi)\right)\phi\, ,
\lb{6.1}
\ee
where the dilaton potential takes the form
\be
V(\varphi)=m^2 (-2\varphi+\varphi^2)\, .
\lb{6.2}
\ee
Integrating over $\phi$ we obtain an effective action which is functional of the metric and the dilaton $\varphi$. It can be evaluated by using the heat kernel technique,
\be
W_{eff}=-\frac{1}{2}\int_{\epsilon^2}^\infty \frac{ds}{s} \Tr K(s,x,x) e^{-sm^2}\, .
\lb{6.3}
\ee
Here $K(s,x,x')$ is the heat kernel for the operator $\left(-\Box +\frac{1}{6}R+V(\varphi)\right)$.  What we shall need is a decomposition of the heat kernel in the powers of the potential
$V(\varphi)$, where   we want to include  the mixed terms linear in the curvature. The required structure of the heat kernel can be extracted from the exhaustive study presented in \cite{Barvinsky:1993en}.  We use the  equations (4.47)-(4.49) in Ref.\cite{Barvinsky:1993en}   and focus only on the terms polynomial in $V$,
that contain maximum  the first power of the curvature and the second derivative of the potential $V$. Among  plenty of terms obtained in \cite{Barvinsky:1993en} we shall need only  the following five terms\footnote{Note that our potential $V$ is related to $P$, used in \cite{Barvinsky:1993en}, as $V=-P$.} 
\be
&&\Tr K(s,x,x)=\frac{s^3}{(4\pi s)^2}\int [(\frac{1}{12}V\Box V-\frac{1}{180}V\Box R) \nonumber \\
&&-s(\frac{1}{24}\Box V V^2
-\frac{1}{720}V^2\Box R-\frac{1}{180}VR^{\mu\nu}\nabla_\mu\nabla_\nu V))]\, .
\lb{6.4}
\ee
The terms  with higher powers of $s$ contain either  higher powers of  potential $V$, higher powers of the curvature, or higher derivatives of $V$.  They are not of our interest here.
The effective action then, after the integration over proper time $s$, the integration by parts and using the Bianchi identities, is a sum of two terms,
\be
&&W_{eff}=W_0+W_1\, ,\nonumber \\
&&W_0=-\frac{1}{384\pi^2}\int(\frac{1}{m^2}V\Box V-\frac{1}{2m^4}V^2\Box V)\,  ,\nonumber \\
&&W_1=\frac{1}{57650\pi^2}\int(\frac{1}{m^4}G^{\mu\nu}\partial_\mu V\partial_\nu V+R(\frac{1}{m^2}\Box V-\frac{1}{2m^4}(\nabla V)^2-\frac{1}{m^4}V\Box V))\, ,
\lb{6.5}
\ee
where $G^{\mu\nu}=R^{\mu\nu}-\frac{1}{2}Rg^{\mu\nu}$ is the Einstein tensor. For the potential $V$, expressed in terms of the dilaton as in (\ref{6.2}), we find
\be
W_1(\varphi)=\frac{1}{57650\pi^2}\int(4(\varphi-1)^2G^{\mu\nu}\partial_\mu\varphi\partial_\nu\varphi -2R(\varphi-1)^3\Box\varphi)\, .
\lb{6.6}
\ee
On-shell, (\ref{2.3}), the second term in (\ref{6.6}) vanishes. For small values of $\varphi$ we then obtain
\be
W_1(\varphi)=\frac{1}{1440\pi^2}\int G^{\mu\nu}\partial_\mu\varphi\partial_\nu\varphi
\lb{6.7}
\ee
in a complete agreement with our expectations. We see that the two-derivatives term $\partial_\mu\varphi\partial_\nu\varphi$ does couple to the Einstein tensor.
The contribution to the entanglement entropy then is equal to
\be
S(\varphi)=\frac{1}{720\pi}\int_\Sigma (\partial_\Sigma\varphi)^2\, .
\lb{6.8}
\ee
For  a scalar field $a_{UV}=1/90\pi^2$. Therefore, by comparing (\ref{6.8}) with (\ref{4.3}), we find that 
in the case of a free field the RG flow  results in an IR theory which does not contain any degrees of freedom, $a_{IR}=0$. This is as expected, see the discussions in \cite{Cappelli:1990yc} and \cite{Komargodski:2011vj}.

The entanglement entropy for the field theory (\ref{6.1}) in flat spacetime should not depend on the value of the non-minimal coupling. As a consistency check on
our analysis let us see whether the result (\ref{6.7}), (\ref{6.8}) is sensitive to the exact value of the non-minimal coupling in (\ref{6.1}).  For this let us shift the 
potential $V\rightarrow V+\xi R$. The modification of the action $W_1$ then comes from a deformation of the action $W_0$ in (\ref{6.5}),
\be
\Delta_\xi W_1=W_0(V+\xi R)-W_0(V)=-\frac{1}{192\pi^2}\int \xi R\left(\frac{1}{m^2}\Box V-\frac{1}{m^4}V\Box V-\frac{1}{2m^4}(\nabla V)^2\right)\, .
\lb{6.9}
\ee
For the potential $V$ in the form (\ref{6.2}) this deformation reads
\be
\Delta_\xi W_1=-\frac{1}{48\pi^2}\int \xi R\left((\varphi-1)(1+2\varphi-\varphi^2)\
\Box\varphi-2\varphi(\varphi-2)(\nabla\varphi)^2\right)\, .
\lb{6.10}
\ee
The first term here vanishes on-shell while the second gives a contribution only in cubic order.  We conclude that our result (\ref{6.7}) is robust and
does not change under a deformation of the non-minimal coupling. So that the entropy (\ref{6.8}) is indeed  not sensitive to the values of the non-minimal coupling in the field action (\ref{6.1}), when evaluated in flat spacetime. In particular, this result should be  valid for a minimally coupled scalar field.

\subsection{Entanglement entropy in  general case, the a-theorem}
%\section{ Remarks on the RG flows in $D=6$}
%\setcounter{equation}0
The analysis which we now perform is inspired by the one made by Komargodski \cite{Komargodski:2011xv} in the two-dimensional case for the c-theorem.
The dilaton $\varphi(x)$   couples to the trace of stress-energy tensor as $\varphi(x)T(x)$. Focusing on terms quadratic in $\varphi$ we find that
\be
\left\langle e^{\int \varphi(x)T(x)}\right\rangle= 1+\frac{1}{2}\int\int \varphi(x)\varphi(y)\left\langle T(x)T(y)\right\rangle+..
=e^{-W(\varphi)}\, ,
\lb{6.11}
\ee
where, in quadratic order,    the  dilaton action is
\be
W(\varphi)=-\frac{1}{2}\int\int \varphi(x)\varphi(y)\left\langle T(x)T(y)\right\rangle\, .
\lb{6.12}
\ee
Following \cite{Cappelli:1990yc} we shall use a spectral representation for the correlation function\footnote{We note that our normalization for the stress energy tensor is
different from the one used in \cite{Cappelli:1990yc}. The zero-spin spectral density $C^{(0)}(\mu)$ is, however, identical to the one used in \cite{Cappelli:1990yc}.}
\be
\left\langle T(x),T(y)\right\rangle =\frac{1}{480\pi^2}\int_0^\infty d\mu \, C^{(0)}(\mu)\mu^4G_\mu(x,y)\, .
\lb{6.13}
\ee
$G_\mu(x,y)$ is  Green's function of a scalar field of mass $\mu$. We extend the relation (\ref{6.13}) to a curved background\footnote{There is an ambiguity in extending  the relevant scalar field operator to a curved metric: one has to specify the value of the non-minimal coupling in the field operator. However, as we discussed this above, the interesting to us part in the dilaton action/entropy does not depend on the the concrete value of this coupling.} .
The dilaton $\varphi(y)$ can be now decomposed  in a covariant Taylor expansion \cite{Petrov}
\be
\varphi(y)=\varphi(x)+\nabla_\alpha\varphi(x)(y-x)^\alpha+\frac{1}{2}\nabla_\alpha\nabla_\beta \varphi(x)\,  (y-x)^\alpha(y-x)^\beta+..\, ,
\lb{6.14}
\ee
valid in the normal coordinates with the origin at the point $x$. In the normal coordinates we evaluate the integral
\be
\int d^4 y\sqrt{g(y)}\, G_\mu(x,y)(y-x)^\alpha(y-x)^\beta=\frac{H^{\alpha\beta}(x)}{\mu^6}\, ,
\lb{6.15}
\ee
where $H^{\alpha\beta}(x)$ is linear in curvature and we omit all other  powers of the curvature. 

In the dilaton action (\ref{6.12}) we are interested in a term linear in the curvature. This term takes the form
\be
W(\varphi)=-\frac{1}{1920\pi^2}\int_0^\infty d\mu\, \frac{C^{(0)}(\mu)}{\mu^2}\int \varphi(x) H^{\alpha\beta}(x)\nabla_\alpha\nabla_\beta\varphi(x)\, .
\lb{6.16}
\ee
This is a general formula valid for any four-dimensional unitary theory. In particular, it can be applied in the case of a free scalar field of mass $m$.
The form of the spectral function $C^{(0)}(\mu)$ in this case is found in  \cite{Cappelli:1990yc}. On the other hand, for a massive fee field we should reproduce (\ref{6.7}).
The result (\ref{6.7}) does not depend on mass $m$ of the field. In particular, it stays the same if $m$ is taken to zero. In this limit
$C^{(0)}(\mu)/\mu^2\rightarrow \delta(\mu)$, as was shown in \cite{Cappelli:1990yc}.  These reasonings help us to  quickly\footnote{ A slower way would be directly compute the integral (\ref{6.15}) using the normal coordinates.}  determine $H^{\alpha\beta}$,
\be
H^{\alpha\beta}(x)=\frac{4}{3}G^{\alpha\beta}(x)\, .
\lb{6.17}
\ee
So that, in the case of a generic quantum field theory  we have 
\be
W(\varphi)=\left[\frac{1}{1440\pi^2}\int_0^\infty d\mu\, \frac{C^{(0)}(\mu)}{\mu^2}\right]\, \int  G^{\alpha\beta}(x)\partial_\alpha\varphi(x)\partial_\beta\varphi(x)\, .
\lb{6.18}
\ee
The dilaton dependence of the entanglement entropy then can be deduced from this action by applying (\ref{singular-curvature}),
\be
S(\varphi)=\left[\frac{1}{720\pi}\int_0^\infty d\mu\, \frac{C^{(0)}(\mu)}{\mu^2}\right] \int_\Sigma (\partial_\Sigma\varphi)^2\, .
\lb{6.19}
\ee
Comparing this with (\ref{4.3}) we find
\be
a_{UV}-a_{IR}=\frac{1}{90\pi^2}\int_0^\infty d\mu\, \frac{C^{(0)}(\mu)}{\mu^2}\, 
\lb{6.20}
\ee
for the difference in the $a$-charges between the UV and IR conformal theories. 
In a unitary theory the spectral density $C^{(0)}(\mu)\geq 0$ and, thus, eq.(\ref{6.20}) implies the a-theorem.

\subsection{Remarks}

Here we make some remarks.

\bigskip

\noindent{\bf 1.}  According to conventional point of view  the $a$-anomaly  shows up  in the correlation functions of the stress-energy tensor in Minkowski spacetime starting with the 3-point functions. We, however, want to stress that in the presence of the conical defects the $a$-anomaly makes its appearance already in the 2-point function.  This observation  explains why, as we  advocate in this paper, the 2-point function of stress-energy tensor should be relevant to the $a$-theorem despite the fact that  in the proof of  \cite{Komargodski:2011vj}, \cite{KST}
it is related to 3- and 4-point functions.
To illustrate our point we notice that by varying (\ref{1}) with respect to metric and using the relations (\ref{singular-curvature}) for the curvature of conical defect we obtain for the $a$-anomaly part 
of the 2-point function
\be
\left\langle T^{\mu\nu}(x)T(y)\right\rangle_{\rm CFT}=2\pi  (1-\alpha) ~a\,\delta_\Sigma\, (\gamma^{\mu\alpha}\gamma^{\nu\beta}-\gamma^{\mu\nu}\gamma^{\alpha\beta})\partial_\alpha\partial_\beta\, \delta(x-y)\, ,
\lb{r1}
\ee
where $\gamma^{\mu\nu}=g^{\mu\nu}-(n^\mu n^\nu)$ is the induced metric on the surface $\Sigma$ and $\delta_\Sigma$ is the delta-function at the surface $\Sigma$.  We see that this correlation function is purely intrinsic, i.e. it involves only derivatives
along the surface $\Sigma$. For the trace we find
\be
\left\langle T(x)T(y)\right\rangle_{\rm CFT}=-2\pi  (1-\alpha) ~a  \, \delta_\Sigma  \, \partial^2_\Sigma\delta(x-y)\, .
\lb{r2}
\ee
Again, we see that this correlation function is purely two-dimensional.

\medskip

\noindent{\bf 2.} The 2-point function (\ref{6.13}) is supposed to be a correlation function on a curved background. Thus, at least in principle, it may contain  information
on $n$-point functions in Minkowski spacetime. On the other hand, we are interested in a very special type of curved background, flat spacetime with a conical defect.
In order to get a 2-point function on this background we may simply start with a 2-point function in Minkowski spacetime and take a certain sum over images. The latter then 
reduces to the Sommerfeld formula (\ref{5.8})  outlined in section 4.3. These reasonings explain our choice for the form of the correlation function (\ref{6.13}) that is a direct
generalization of the respective expression, obtained in \cite{Cappelli:1990yc}, valid in Minkowski spacetime with same spectral function $C^{(0)}(\mu)$.

\medskip

\noindent{\bf 3.} It is possible that (\ref{6.18})-(\ref{6.20}) are not general enough and should be supplemented by  some other contributions when extended to higher spins.
Indeed, if applied to   free fermions\footnote{We thank the referee for this observation.} these formulas give $\Delta a_f=6 \Delta a_s$ for the  difference in the $a$-anomaly relative to the scalar case.
The actual fermion/scalar ratio, however, should be  11. A possible source of the discrepancy is that there may be an extra contact term, similar to the one in (\ref{r2}), present in the correlation function (\ref{6.13}) when considered on a conical background. This term would not  correspond to propagating states and hence  not be visible in the spectral  
representation of \cite{Cappelli:1990yc}.
The appearance of such a  contact term, however, does not look natural from the point of view of the Sommerfeld formula.  It would be interesting to understand better the
modifications, if any, of (\ref{6.13}) for higher spins and the relations  to higher point correlation functions in Minkowski spacetime in order to make a more direct contact with the proof of \cite{Komargodski:2011vj}.

\section{ Does the a-theorem follow from the c-theorem?}
\setcounter{equation}0
Let us choose  the entangling surface to satisfy the condition $K_\Sigma=0$ so that, in a generic 4d CFT, the surface conformal anomaly  is entirely due to intrinsic geometry of the surface.
The anomaly part of the entanglement entropy (\ref{28})
\be
S=\frac{a\pi}{ 32}\int_\Sigma R_\Sigma \frac{1}{\Delta_\Sigma} R_\Sigma
\lb{7.1}
\ee
then is identical to the anomaly part in the effective action in a two-dimensional conformal field theory. This observation suggests that there might be a closer relation between the two conformal theories. In particular,  the $c$-charge in two dimensions and the $a$-charge in four dimensions are related as follows
\be
c=3\pi^2 a\, .
\lb{7.2}
\ee
In two dimensions, if the theory is unitary, the $c$-charge is positive. 
The relation (\ref{7.2}) then implies that  the $a$-central charge is a positive quantity, $a>0$. 

Same relation (\ref{7.2}) is valid for the RG flows of the central charges, $\Delta c=3\pi^2\Delta a$, as is seen from (\ref{4.2}), (\ref{4.3}).
This can be generalized to higher dimensions. In $d=6$ the $a$-anomaly part of the dilaton action is
\be
W_{d=6}(\tau)=-\Delta a_6\int_{M_6}(\tau E_6+..)\, ,
\lb{7.3}
\ee
where $E_6$ is the Euler density in six dimensions, $\frac{1}{384\pi^3}\int_{M_6}E_6$ is the Euler number, and the terms of higher powers in $\tau$ are obtained by the conformal symmetry, see \cite{Elvang:2012st} for a relevant discussion. Applying (\ref{singular-curvature}) to (\ref{7.3}) one finds the dilaton entropy
for an  entangling co-dimension two surface,
\be
S_{d=4}(\tau)=6\pi \Delta a_6\int_{\Sigma_4}(\tau E_4+..)\, .
\lb{7.4}
\ee
(We note that the Euler number of $\Sigma_4$ is $\frac{1}{32\pi^2}\int_{\Sigma_4}E_4$.)
In this formula we used the relation found in \cite{Fursaev:1995ef}  for the Euler densities in the bulk geometry and in the singular surface. 
The higher powers of the dilaton field in (\ref{7.4}) are fixed by the conformal symmetry on the surface. Since this conformal extension is unique 
the dilaton entropy (\ref{7.4}) is precisely the dilaton action (\ref{2.4}) in $d=4$. The relation between the RG flows of  $a$-charges in $d=4$ and $d=6$ (in the normalization which we use here)
is
\be
\Delta a_4=384\pi \Delta a_6\, .
\lb{7.5}
\ee
Clearly, this can be extended to any higher dimension.

We have shown in this paper that  the a-theorem in $d=4$ is due to positivity of the 2-point correlation function of stress-energy tensor. This is similar to the proof of the c-theorem originally presented in \cite{Cappelli:1990yc} (see also discussion in \cite{Komargodski:2011xv}). Both theorems, thus, involve the same properties of the correlation functions. Moreover, for the entanglement entropy (\ref{7.1}) the statement on decreasing of charge $a$ appears to have the form of a $c$-theorem.
These observations suggest that there might be a deeper relation between the two theorems and, possibly, the a-theorem may simply follow from the c-theorem. The c-theorem should be properly applied to  the entanglement entropy of a four-dimensional theory. What we could gain from this relation, provided it is better established, is the complete extension of the properties of the RG flows in two dimensions to the four-dimensional field theories. Most importantly, it might be possible to  demonstrate, using this similarity, that  the RG flow in four dimensions is a gradient flow. We do not develop this direction here leaving it to a future study.

\section{ Conclusions}
\setcounter{equation}0

We have reformulated the a-theorem in terms of the entanglement entropy. Our approach is motivated by the recent works of Komargodski and Schwimmer \cite{Komargodski:2011vj}, \cite{Komargodski:2011xv}, where they suggested to analyze the RG flows of the dilaton action. We applied this idea to the study of the RG flows of the entanglement entropy 
of a generic four-dimensional CFT. It was found that the quartic term in the dilaton action, $(\partial\tau)^4$, which was important for the analysis in \cite{Komargodski:2011vj}, \cite{Komargodski:2011xv} does not contribute at all to the entropy. Instead, the important contribution comes from the term $G^{\mu\nu}\partial_\mu\tau\partial_\nu\tau$,
where $G^{\mu\nu}$ is the Einstein tensor.  This term in the dilaton action is quadratic in $\tau$ and, thus, can be related to a 2-point function of the trace of  stress-energy
tensor. Modulo the technical details this is exactly as in the demonstration of the c-theorem in two dimensions, \cite{Cappelli:1990yc}, \cite{Komargodski:2011xv}.
The 2-point functions and their positivity properties  are easier to analyze than that of the 4-point functions effectively    used in the approach of \cite{Komargodski:2011vj}. 
Our proof of the a-theorem then goes along the lines similar to the proof of the c-theorem. 

We propose that this similarity may have some deeper reasons and, in fact, the a-theorem
may be a consequence of the c-theorem. This conjecture is based on the following observations: 1) the a-anomaly part in the entanglement entropy in a generic 4d CFT is identical to the effective action of a two-dimensional CFT defined on the entangling surface; 2) in the analysis of the RG flows, the dilaton part of the entanglement entropy in a four-dimensional theory is identical to the dilaton action defined on the co-dimension two entangling surface; 3) the proof of the a-theorem for the entanglement entropy involves the same positivity property of the spectral representation of  the 2-point function of 
stress-energy tensor as in the c-theorem. The points 1) and 2) can be extended to higher dimensions and involve a certain relation between the $a$-charges in dimensions $d=2k+2$ and $d=2k$. It would be interesting to understand better  all these relations.

\section*{Acknowledgements} 

I would like to thank Z. Komargodski  and A. Schwimmer for correspondence.  I thank Yu. Nakayama and A. Cappelli for useful discussions.

\end{document}